\documentclass[pdflatex,sn-mathphys]{sn-jnl}% Math and Physical Sciences Reference Style
\usepackage{graphicx}
\usepackage{subfigure}
\jyear{2021}%

\theoremstyle{thmstyleone}%
\theoremstyle{thmstyletwo}%

\theoremstyle{thmstylethree}%

\begin{document}

\title[An improved two-threshold  quantum segmentation algorithm for NEQR image]{An improved two-threshold  quantum segmentation algorithm for NEQR image }

%%=============================================================%%
%% Prefix	-> \pfx{Dr}
%% GivenName	-> \fnm{Joergen W.}
%% Particle	-> \spfx{van der} -> surname prefix
%% FamilyName	-> \sur{Ploeg}
%% Suffix	-> \sfx{IV}
%% NatureName	-> \tanm{Poet Laureate} -> Title after name
%% Degrees	-> \dgr{MSc, PhD}
%% \author*[1,2]{\pfx{Dr} \fnm{Joergen W.} \spfx{van der} \sur{Ploeg} \sfx{IV} \tanm{Poet Laureate} 
%%                 \dgr{MSc, PhD}}\email{iauthor@gmail.com}
%%=============================================================%%
\author[1]{\fnm{Lu} \sur{Wang}}\email{Lu\_Wang\_MT@163.com}

\author[1]{\fnm{Zhiliang} \sur{Deng}}\email{mtsdzl@163.com}

\author*[2]{\fnm{Wenjie} \sur{Liu}}\email{wenjiel@163.com}

\affil[1]{\orgdiv{School of Automation}, \orgname{Nanjing University of Information Science  and Technology}, \orgaddress{ \city{Nanjing}, \postcode{210044}, \state{Jiangsu}, \country{China}}}

\affil*[2]{\orgdiv{School of Computer and Software}, \orgname{Nanjing University of Information Science  and Technology},
\orgaddress{ \city{Nanjing}, \postcode{210044}, \state{Jiangsu} \country{China}}}

%%==================================%%
%% sample for unstructured abstract %%
%%==================================%%

\abstract{The quantum image segmentation algorithm is to divide a quantum image into several parts, but most of the existing algorithms use more quantum resource(qubit) or cannot  process the  complex image. In this paper, an improved two-threshold  quantum segmentation algorithm for NEQR image is proposed, which can  segment the complex gray-scale image into a clear ternary image  by using fewer qubits and can be scaled to use $n$ thresholds for $n+1$ segmentations. In addition, a feasible quantum comparator  is designed to distinguish the gray-scale values with  two thresholds, and then a  scalable 
quantum circuit is designed to segment the  NEQR image.  For a  ${2^n} \!\times\! {2^n}$  image with $q$ gray-scale levels, the 
quantum cost of our  algorithm can be reduced to 60$q$-6, which is lower than   other existing  quantum algorithms and  dose not increase with the image's size increases. The experiment on  IBM Q   demonstrates that our algorithm can effectively segment the image. }

%%================================%%
%% Sample for structured abstract %%
%%================================%%


%%\pacs[JEL Classification]{D8, H51}

%%\pacs[MSC Classification]{35A01, 65L10, 65L12, 65L20, 65L70}

\maketitle

\section{Introduction}\label{sec1}
Due to the unique parallelism and entanglement characteristics, quantum computing  can achieve quantum acceleration, which makes the calculation speed improved to varying degrees compared with classical computing. Quantum image processing (QIP) is a cross-discipline of quantum computing and image processing, and  this makes it possible to solve the  real-time problem that encountered by the classical counterpart. 

In the past 20 years, the QIP  has developed rapidly, and the specific development process   can be found in Refs \cite{Yan2017,Cai2018}. 
It can be divided into two stages: one is to store classical images in a quantum image representation model, and the other is the study of QIP algorithms. In the first stage, there are mainly: qubit lattice representation \cite{Venegas-Andraca2003}, real ket representation \cite{Latorre2005}, entangled images representation \cite{Venegas-Andraca2010},  flexible representation of quantum image (FRQI) \cite{Le2011}, the multi-channel RGB images representation of quantum images (MCQI) \cite{Sun2013},  and quantum probability image encoding representation (QPIE) \cite{Yao2017},  novel enhanced quantum image representation (NEQR) \cite{Zhang2013}, an improved NEQR (INEQR) \cite{Jiang2015}, a generalized model of NEQR (GNEQR) \cite{Zhang2015L}, a novel quantum representation of color digital images (NCQI) \cite{Sang2017}.  Among them, the NEQR model is widely used due to its simplicity of operation. Based on different quantum image representation models, the corresponding
QIP algorithms also develop rapidly, such as geometrical transformation of quantum image \cite{Zhou2017}, quantum image encryption \cite{Zhou2013}, feature extraction of quantum image \cite{Hancock2015}, quantum image scrambling \cite{Zhou2015}, quantum image morphological
operations \cite{Yuan2015},  quantum image watermarking \cite{Song2014}, quantum image filtering \cite{Li2017}, quantum image steganography \cite{Zhao2021}, quantum image stabilization \cite{Yan2016}, quantum image bilinear interpolation \cite{Yan2021}, quantum image edge detection \cite{Zhou2016,Fan2019,Chetia2021,Liu2022}, quantum image segmentation \cite{Caraiman2014,Caraiman2015,Xia2019}, etc.  Although the research of quantum image processing technology is gradually deepening, it is still in the initial stage as a whole, and the development direction is not balanced. The results of these studies have shown that quantum image processing is exponentially faster than classical image processing, and it saves a lot of computing resources. There are many quantum image processing algorithms that are theoretically feasible, but they are not performed on real quantum computers or simulators, and these algorithms require a large number of qubits, which  is not feasible in this Noisy Intermediate-Scale Quantum (NISQ) era. So, further research is needed for deeper image processing algorithms.

Image segmentation is the first step of image analysis, and it can divide the image into several parts to clarify the target in the image. The most common image segmentation methods are: threshold-based segmentation, region-based segmentation, edge detection-based segmentation, and segmentation combined with specific tools.  In recent years, researchers have correspondingly proposed several quantum image segmentation algorithms. In 2013, Li et al. \cite{Li2013} used a quantum search algorithm  to  segment the quantum image. The theoretical analysis shows that a better effect is achieved, but no specific oracle  was provided, which made the simulation very difficult. In 2014, Caraiman et al. \cite{Caraiman2014} proposed a histograms-based  quantum image segmentation algorithm. They used the quantum computing  performance  to obtain exponential speedup than the classical algorithm, but still no specific oracle circuit is given. A year later, they \cite{Caraiman2015} proposed a quantum image segmentation algorithm based on a single threshold. They gave a specific implementation circuit, but the number of qubits required is too large to be simulated on the existing platform.  In 2019, Xia et al. \cite{Xia2019} proposed a  multi-bit quantum comparator and  applied it in image binarization. They chose a single threshold to perform binary segmentation on quantum images, but the number of qubits required was relatively large. Moreover, the algorithm quantum  cost is relatively high, which is very unsuitable for simulation in the NISQ era. In fact, none of the above algorithms are simulated on a quantum computer or quantum simulator. In 2020, Yuan et al. \cite{Yuan2020} proposed a dual-threshold quantum image segmentation algorithm with a small number of qubits and simulated it on a quantum simulator. However, the algorithm  only segments pixels smaller than the low threshold and larger than the high threshold to 0, and the pixels between the thresholds cannot be processed and remain unchanged, so only at most two segmentations can be made and the target segmentation is not clear, which cannot be scalable and is not suitable for some complex images.  In addition, quantum  cost is still high. In practical use, we need to segment multiple targets in the image according to the situation, which requires the algorithm to have better scalability, which is impossible for the above algorithms.

In order to  increase the practicability of the quantum image segmentation algorithm in this NISQ ear, we have carried out further research in this paper, and the main contributions  are summarized as follows. 
\begin{itemize}
    \item An improved two-threshold  quantum segmentation algorithm for NEQR image is proposed, which can be scaled to use $n$ thresholds for $n+1$ segmentations.
    \item A quantum comparator with lower quantum cost is designed to distinguish the gray-scale values with two thresholds. Then, a complete and scalable quantum segmentation circuit is designed to clearly segment the complex quantum images  by using few qubits.
    \item We  verify  the  superiority  and  feasibility  of  our  proposed  algorithm  by analyzing  the  circuit  quantum cost  and  performing the simulation  experiment on  IBM Q platform \cite{IBM} through Qiskit extension \cite{Qiskit}, respectively.
\end{itemize}
  
The rest of this paper is organized as follows: In Sec. 2, The basics of quantum circuits and the NEQR model are introduced. In Sec. 3, we first introduce the two-threshold quantum image segmentation algorithm, and then,   a quantum comparator and a complete segmentation  circuit are designed in detail.   In Sec. 4,   the circuit quantum cost is analyzed and the simulation experiment is  performed on the IBM Q platform. Finally, the discussion is given in Sec. 5.

\section{Related Works}\label{sec2}
\subsection{Quantum circuit}
Quantum circuit is to perform a series of logical  operations on n qubits, and finally perform a measurement, which  is an implementation form of quantum computing. A quantum circuit represents the sequence of time from left to right, and each line represents one qubit. There are many logic gates called quantum gates in the circuit,  which are the basis of quantum circuits and represent the unitary operation of qubits. Unlike traditional logic gates, quantum logic gates are reversible, and traditional calculations can  be represented by reversible gates. This means that quantum circuits can simulate the operation of all traditional circuits. Combining some basic quantum gates can form  more complex quantum gates, and these complex quantum gates can act on multiple qubits at the same time. Therefore, a circuit composed of  multiple quantum gates is a quantum circuit.  Common quantum gates include Hadamard gate (H gate), Pali-X gate (X gate), Control-NOT gate (CNOT gate) and Toffoli gate (CCNOT gate). The symbol and unitary matrix of each quantum gate are shown in Table 1.

\begin{figure}
 \flushleft
    \includegraphics[width=9cm]{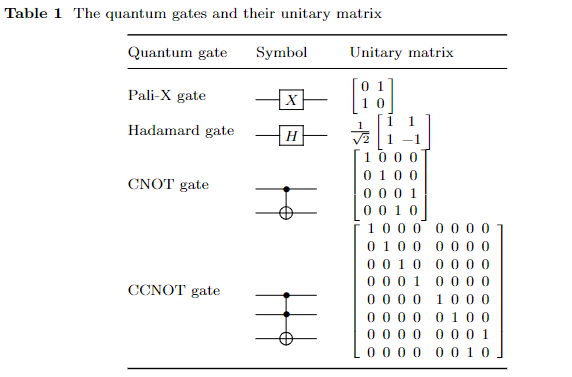 }
    %\caption{An example of a 2×2 image}
    \label{tabl}
\end{figure}

%%  \Qcircuit @C=1em @R=0.7em {   
%%  & \gate{\lvert0\rangle} & \qw }

\subsection{NEQR}
A digital image is composed of pixels, and the pixels contain position information and color information. The NEQR model uses three entangled qubit sequences to store these two kinds of information, and the whole image is stored  in the superposition of the two qubit sequences. Assuming the image size is $2^n\times2^n$,  we need to use two entangled qubit sequences of length $n$  to represent the position information x-axis and y-axis coordinates. For general gray-scale images, the gray-scale value range is $[0,2^q-1]$, so a $q$-length qubit sequence is needed to store the gray-scale value of the pixels. The entire model is the tensor product of these three entangled qubit sequences, so that all pixels can be stored and processed at the same time. Then the NEQR model of the quantum image can be written in the form of the quantum superposition state shown in Eq. (1) \cite{Zhang2013}.
\begin{equation}
   \lvert {\rm{I}} \rangle  = \frac{1}{{{2^n}}}\sum\limits_{Y = 0}^{{2^n} - 1} {\sum\limits_{X = 0}^{{2^n} - 1} {\lvert {{C_{YX}}} \rangle  \otimes \lvert Y \rangle \lvert X \rangle } }  = \frac{1}{{{2^n}}}\sum\limits_{YX = 0}^{{2^{2n}} - 1} {\mathop  \otimes \limits_{k = 0}^{q - 1}\lvert {C^K_{YX}} \rangle \mathop  \otimes \limits_{}^{} \lvert {YX} \rangle  } 
\end{equation}
where  $\lvert {{C_{YX}}} \rangle  = \lvert {C_{YX}^{q - 1},C_{YX}^{q - 2},{ \cdots ^{}}C_{YX}^{1}C_{YX}^{0}} \rangle$ represents the quantum image gray-scale values, $C_{YX}^k \in \left\{ {0,1} \right\}$, $k = q - 1,q - 2, \cdots ,0$. $  \lvert {YX} \rangle  = \lvert Y \rangle \lvert X\rangle  = \lvert {{Y_{n - 1}},{Y_{n - 2}}, \cdots {Y_0}}\rangle \lvert {{X_{n - 1}},{X_{n - 2}}, \cdots {X_0}} \rangle $ represents the 
position of the pixel in a quantum image, ${Y_t},{X_t} \in \left\{ {0,1} \right\}$.
 
\begin{figure}
 \centering
    \includegraphics[width=3cm]{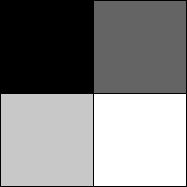 }
    \caption{An example of a 2×2 image}
    \label{Figl}
\end{figure}

Fig. \ref{Figl} shows an example of a grayscale image of size 2×2, and the corresponding NEQR expression of which is given as follows

\[\begin{array}{l}
\lvert I \rangle  = \frac{1}{2}\left( {\lvert 0 \rangle \lvert{00} \rangle  + \lvert {100} \rangle \lvert {01} \rangle  + \lvert {200} \rangle \lvert {10} \rangle  + \lvert{255} \rangle \lvert {11} \rangle } \right)
\end{array}\]
$$\begin{array}{l}
\begin{array}{*{20}{c}}
 = 
\end{array}\frac{1}{2}\left( \begin{array}{l}
\lvert {00000000} \rangle \lvert{00} \rangle  + \lvert {01100100} \rangle \lvert {01} \rangle \\
 + \lvert {11001000}\rangle \lvert{10} \rangle  + \lvert {11111111} \rangle \lvert {11} \rangle 
\end{array} \right)
\end{array}
\eqno{(2)}$$

\section{The improved two-threshold quantum image segmentation algorithm}\label{sec3}
In this section, we first introduce the two-threshold quantum image segmentation algorithm,  then, a quantum comparator with high performance is designed to compare the thresholds and the gray-scale values, and a complete quantum circuit with high parallelism is designed to segment the NEQR image. 

\subsection{The quantum image segmentation algorithm }\label{subsec2}
 After the quantum image preparation is complete, a high threshold and a low threshold are given. Then, we set the gray-scale value of the pixels  smaller than the low threshold to $g_1$,  set the gray-scale value of the pixels  between the high  threshold and the low threshold  to $g_2$, and set the gray-scale value of the pixels  larger than the high threshold  to $g_3$. In this way, the image is 
segmented to a ternary image, which can achieve the purpose of clearly segmenting the image.

 Supposing $T_H$ is the high threshold, $T_L$ is the low threshold, $f(x,y)$ is the original image, and $g(x,y)$ is the corresponding  segmented image. So, the description of the segmentation algorithm is as follows:
\[ g(x,y)=\left\{
\begin{array}{rcl}
g_1,       &      & {f(x,y)     <      T_L}\\
g_2,     &      & {T_L \leq f(x,y)  < T_H}\\
g_3,    &      & {f(x,y) \ge T_H}
\end{array} \right.\eqno{(3)} \]

In practical applications, the image may contain more targets, which requires us to segment the image multiple times to achieve the purpose of clear segmentation. So we need to set multiple thresholds to divide pixels into multiple different classes. Then the pixels between each two thresholds are of a class, and the double-threshold segmentation can be scaled to multi-threshold segmentation. The formula is described as follows.

\[ g(x,y)=\left\{
\begin{array}{rcl}
g_1,       &      & {f(x,y)     <      T_1}\\
g_2,     &      & {T_1 \leq f(x,y)  < T_2}\\
g_3,     &      & {T_2 \leq f(x,y)  < T_3}\\
\vdots    &      & {\vdots }\\
g_n,     &      & {T_{n-1} \leq f(x,y)  < T_n}\\
g_{n+1},     &      & {f(x,y)  \ge T_n}\\
\end{array} \right.\eqno{(4)} \]

Since multi-threshold segmentation is an extension of the improved two-threshold segmentation, in this paper we design the quantum circuits by using two-threshold as an example.
It can be seen from the above formula that some quantum comparators are needed to compare the threshold  with the gray-scale values of the quantum image. To facilitate the introduction of the algorithm, we choose the gray-scale value from 0 to 7 as an example. So, the gray-scale values of the quantum image and the segmented image both need 3 qubits to represent( $\lvert C \rangle  = \lvert {{C_1}} \rangle \lvert {{C_2}} \rangle \lvert {{C_3}} \rangle $), therefore, the threshold  also needs to be represented by 3 qubits. And because the high threshold and low threshold can be compared in sequence, 3 qubits can be shared in this way, so the threshold can be expressed as $\lvert T \rangle  = \lvert {{T_1}} \rangle \lvert {{T_2}} \rangle \lvert {{T_3}} \rangle $. The thresholds  in this paper are 
prepared in advance. The high threshold is set to $\lvert {T_H}\rangle =\lvert 100 \rangle$, and the low threshold is set to  $\lvert {T_L}\rangle =\lvert 010 \rangle$, then, $g_1=000$, $g_2=100$, $g_3=111$.  We choose a pixel of the original image  as an example, such as $\lvert 0010010 \rangle$, where the high three qubits represent the gray-scale value, and the low four qubits represent the position information. Therefore, the segmented result  for this pixel is  $\lvert 0000010 \rangle$. 

\subsection{Quantum segmentation circuit design  }

In the quantum image segmentation circuit, the name of each qubit is marked on the far left, and the default initial state of every qubit is $\lvert 0 \rangle$. The circuit is mainly composed of six parts. The first part is the input of quantum image information, including position information and gray-scale value information. The second part is to compare the gray-scale values of the quantum image with the high threshold, and the third part is to  segment the pixels with the gray-scale values larger than the high threshold in the quantum image. The fourth part is to compare the gray-scale values of the quantum image with the low threshold, and the fifth part is to segment the pixels with the gray-scale values less than the low threshold in the quantum image. The sixth part is to use the result of the previous threshold comparison to segment the pixels whose gray-scale values are between the high and low thresholds. The following will be explained in detail in the order of these 
six parts. 

\begin{figure}
    \centering
   \includegraphics[width=5.5cm]{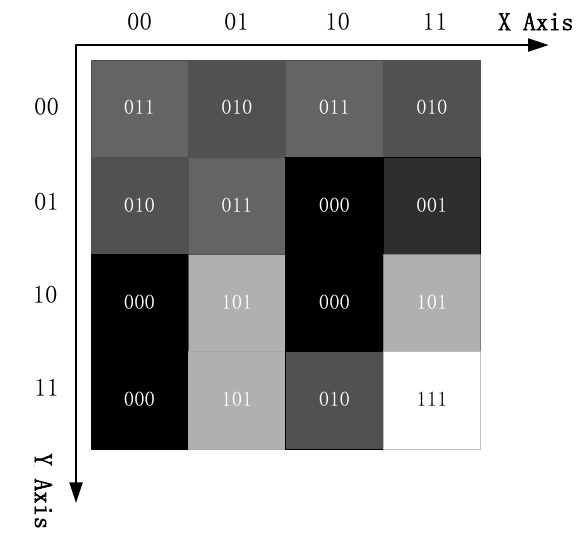 }
    \caption{Schematic of a 4×4  gray-scale image}
    \label{Fig2}
\end{figure}
In order to better describe the algorithm, Fig. \ref{Fig2}  gives a 4×4 image as an example, and the gray-scale value range is [0,7]. The X-Axis and Y-Axis  respectively represent the relative position of each pixel in the image, and each axis requires  two qubits to store. The gray-scale value of each pixel is  marked in the image by using three qubits. The NEQR model of the  image is expressed as Eq. (5).

\[\begin{array}{l}
\lvert {I} \rangle  = \frac{1}{{{2^2}}}(\lvert {0110000} \rangle  + \lvert {0100001} \rangle  + \lvert {0110010} \rangle  + \lvert {0100011} \rangle+\\
\begin{array}{*{20}{c}}
{\begin{array}{*{20}{c}}
{}&{\begin{array}{*{20}{c}}
{}&{}&{}&{}&{}&{}&{}
\end{array}}
\end{array}}&{ \lvert {0100100} \rangle  + }
\end{array} \lvert {0110101} \rangle  + \lvert {0000110} \rangle  + \lvert {0010111}\rangle+\\
\begin{array}{*{20}{c}}
{\begin{array}{*{20}{c}}
{}&{\begin{array}{*{20}{c}}
{}&{}&{}&{}&{}&{}&{}
\end{array}}
\end{array}}&{  \lvert {0001011} \rangle + }
\end{array} \lvert {1011001} \rangle  + \lvert {0001010} \rangle + \lvert {1011011}\rangle+ \\
\begin{array}{*{20}{c}}
{\begin{array}{*{20}{c}}
{}&{\begin{array}{*{20}{c}}
{}&{}&{}&{}&{}&{}&{}
\end{array}}
\end{array}}&{ \lvert {0001100} \rangle + }
\end{array} \lvert {1011101} \rangle  + \lvert {1011110} \rangle + \lvert {1111111} \rangle)\\
\end{array}\eqno{(5)}\]

\begin{figure}
    \centering
    \includegraphics[width=6cm]{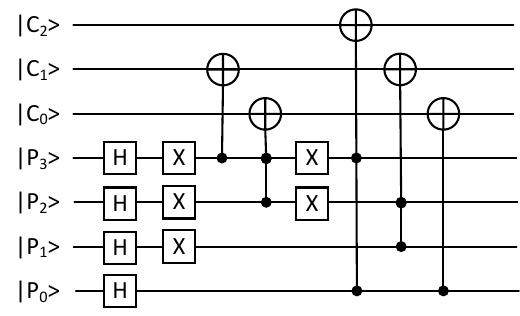 }
    \caption{Quantum image preparation circuit}
    \label{Fig3}
\end{figure}

The first part is  to  input the prepared quantum image into the segmentation circuit. Before this, we have converted the position information and gray-scale value information of the digital image into the form of a quantum image.  Since the qubits' initial state  on the IBM Q platform \cite{IBM} is  $\lvert 0 \rangle$, the quantum circuit needs to be initialized according to our algorithm. Because the initial state of the auxiliary qubits and threshold qubits are some certain values, only ordinary initialization is required. For example, the qubits with the threshold value of $\lvert 010\rangle$, $\lvert 0\rangle$ does not need any change, and only a NOT gate is needed to directly change the second $\lvert 0\rangle$  to $\lvert 1\rangle$. But for quantum images, we do not use ordinary initialization methods. Because the NEQR model uses three entangled qubit sequences to store the image gray-scale value and position information, and  the whole imageis stored in the superposition state of the three qubit sequences, so it is necessary to use the relationship between the pixels' positions and  gray-scale values in a quantum image to initialize the quantum image preparation circuit \cite{Le2011}. It is realized by the combination of H gate, NOT gate and CNOT gate, as shown in Fig. \ref{Fig3}. Among them, $C_2$, $C_1$, $C_0$ represent gray-scale value information, and $P_3$, $P_2$, $P_1$, $P_0$ represent position information. First. By using H gate transformation on the position information qubits, the four qubits  are transformed from $\lvert 0 \rangle$ to a superposition state of $\lvert 0 \rangle$ and $\lvert 1 \rangle$. This makes these four qubits become the superposition state of all  positions of the pixels in the whole image, and the consumption of qubits representing images can be saved through the superposition of qubits. Then  the CNOT gates  are used to realize the entanglement of the position sequence and the gray-scale value sequence. The circuit realization is shown in Fig. \ref{Fig3}. 

\begin{figure}
\centering
    \subfigure[1-qubit comparator circuit]{\includegraphics[width=4cm]{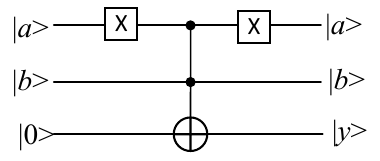 }}\\
    \subfigure[2-qubit comparator circuit]{\includegraphics[width=6cm]{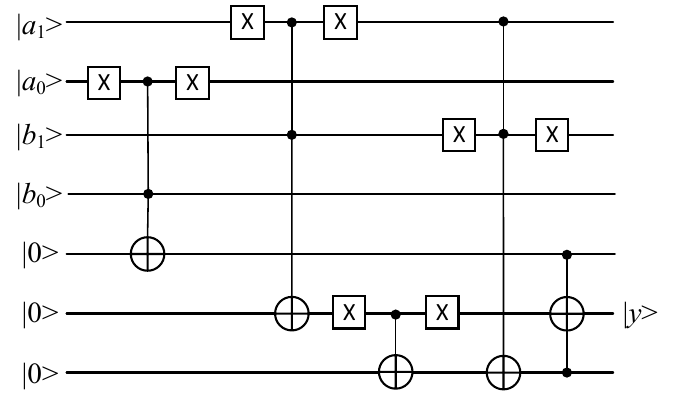 }}\\
    \subfigure[3-qubit comparator circuit]{\includegraphics[width=8cm]{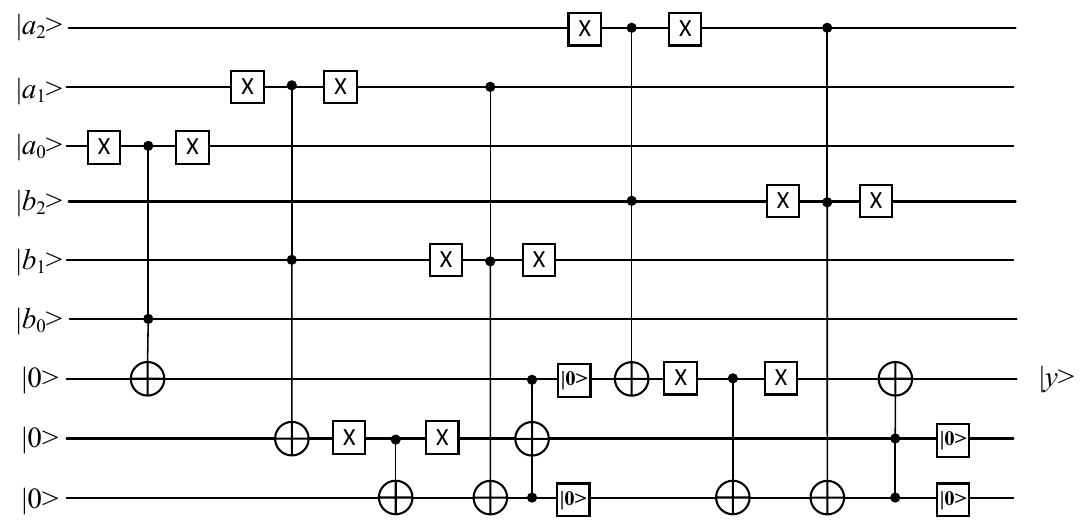 }}
    \caption{Quantum comparison circuit}
    \label{Fig4}
\end{figure}

\begin{figure}
        \centering
        \subfigure[Quantum circuit of comparing one threshold with the whole quantum image]{\includegraphics[width=8cm]{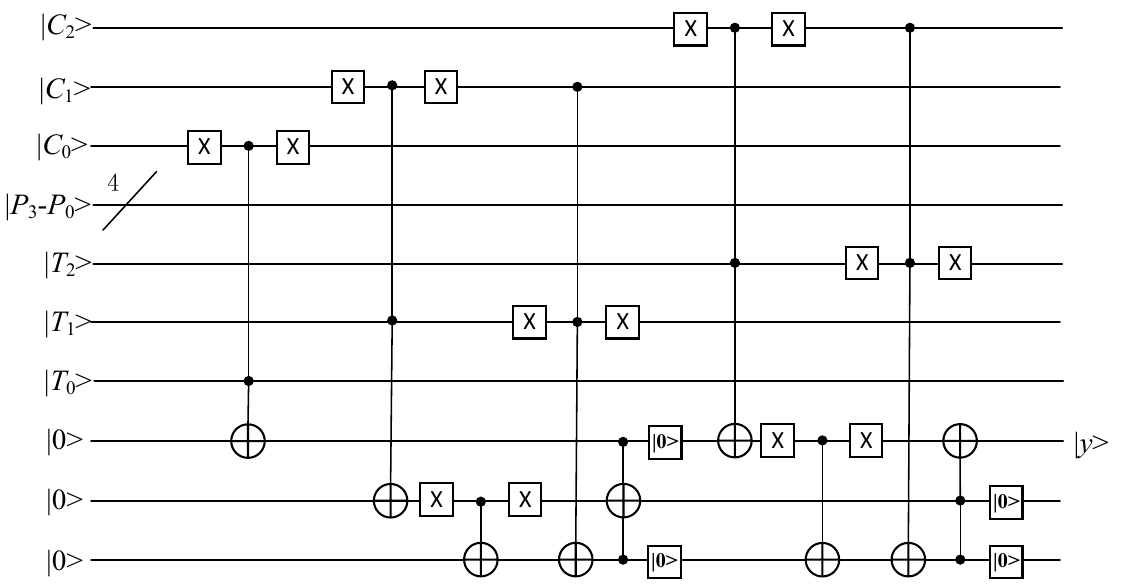 }}
        \subfigure[The quantum circuit diagram of the circuit (a)   ]{\includegraphics[width=5cm]{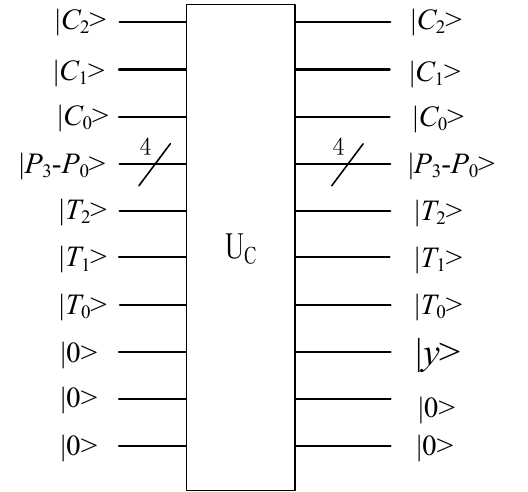 }}
        \caption{The quantum circuit of 3-qubit quantum comparator   and its simplified diagram}
        \label{Fig5}
    \end{figure}

After the quantum image is input, we need to initialize the threshold. Because the threshold is a certain value set in advance, the initialization of the threshold adopts the ordinary initialization method  described above. Then, the quantum  comparator is needed to compare the gray-scale values of the quantum image with the threshold. Inspired by quantum bit string comparator (QBSC) \cite{Oliveira2007}, we design a more suitable comparator in this paper. The 1-qubit quantum comparator is the basis of quantum comparator, as shown in Fig. \ref{Fig4}(a). The circuit takes $\lvert a\rangle$ and $\lvert b\rangle$ as input,   $\lvert y\rangle$ as output. If $a\ge b$, then $ y=0$; if $a<b$, then $y=1$. The 1-qubit comparator only needs to perform one comparison, and the circuit requires a total of three qubits. Two comparison qubits are used to compare the relationship between $\lvert a\rangle$ and $\lvert b\rangle$, and one auxiliary qubit is used to store the comparison result. The 2-qubit comparator needs to perform twice comparisons, so it needs to add one qubits on the basis of 1-qubit comparator to store the result of the second comparison. In addition, another one qubit is needed to compare the two results of the twice comparisons. At this time, if $a_1b_1=01$, it means $a<b$ and it directly outputs the comparison result $y=1$. If $a_1b_1=00$ or 11, the CNOT gate and Toffoli gate are required to use the comparison result of $a_0b_0$ output  $y=0$ or 1. If $a_1b_1=10$, it means $a>b$, a Toffoli gate is required to make the output result $y=0.$ Therefore,  the final comparison result is obtained. As shown in Fig. \ref{Fig4}(b), this circuit requires a total of 7 qubits, among them, 4 qubits are used for comparison,  1 qubits are used to store the comparison result, and 2 qubits are used to form redundant information qubits. When the number of qubits that need to be compared continues to increase, we just need to use  reset operation \cite{Shende2005} to reset the qubits that constitute the redundant information of the circuit in the 2-qubit comparator, and we can continue to compare the increased qubits without adding more  auxiliary qubits. In this paper, a 3-qubit comparator is needed to compare the magnitude between the gray-scale values of the quantum image and the threshold. This need to perform another comparison based on the comparison result of the first two qubits on the basis of the 2-qubit comparator and finally $\lvert y\rangle$ is output.  The circuit is shown in Fig. \ref{Fig4}(c). According to  $\lvert y\rangle$, the magnitude relationship between the gray-scale value of the pixel and the threshold can be obtained.  The complete 3-qubit comparison circuit  of comparing one threshold with the whole quantum image is shown in Fig. \ref{Fig5}(a). Where $\lvert {C_2C_1C_0}\rangle$ encodes the gray-scale value information, $\lvert P_3P_2P_1P_0\rangle$ encodes the position information, $\lvert T_2T_1T_0\rangle$ encodes the threshold information, and the remaining qubits are auxiliary qubits. Before image segmentation, the gray-scale values of all pixels in the prepared quantum image must be compared with the threshold, and finally the every piexls' comparison results  $\lvert y\rangle$ are obtained. Then, the threshold qubits and the other two auxiliary qubits are all set to $\lvert 0\rangle$, and then we can continue to use these qubits during next comparison. In order to facilitate the use of the 3-qubit comparison circuit later, we have given a simplified form of the circuit, as shown in Fig. \ref{Fig5}(b). 
    
\begin{figure}
        \centering
        \subfigure[]{\includegraphics[width=7cm]{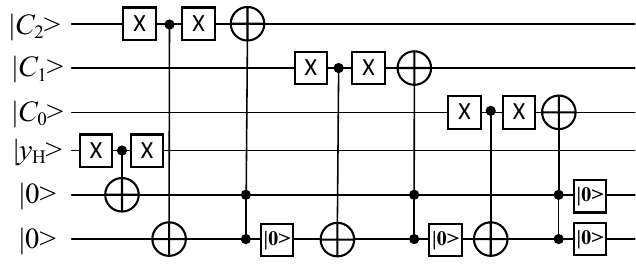 }}
        \subfigure[]{\includegraphics[width=2.5cm]{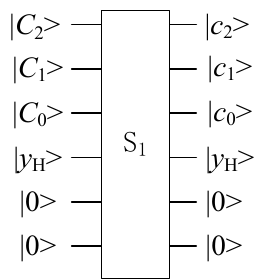 }}
        \caption{The segmentation circuit diagram of pixels larger than the high threshold and its simplified diagram ($S_1)$}
        \label{Fig6}
    \end{figure}

    After the first threshold comparison, the pixels that are larger than the high threshold  are segmented according to the comparison result, and their gray-scale values are set to 111. Namely if $y_H=0$, the gray-scale value information qubits are transformed to 111 one by one. When the segmentation operation of one gray-scale value qubit is completed,  the auxiliary qubit needs to be set to $\lvert 0\rangle$, and the next gray-scale value qubit can continue to be transformed, which can save the number of qubits. As shown in Fig. \ref{Fig6}.
    
\begin{figure}
        \centering
        \subfigure[]{\includegraphics[width=6cm]{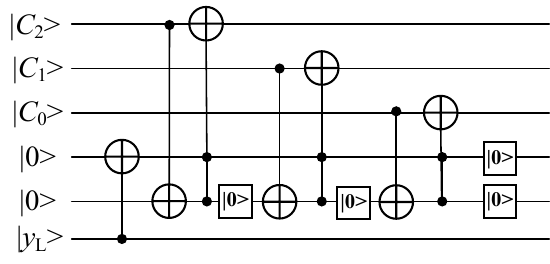 }}
        \subfigure[]{\includegraphics[width=2.5cm]{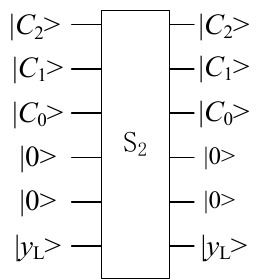 }}
        \caption{The segmentation circuit of pixels smaller than the low threshold and its simplified diagram $(S_2)$}
        \label{Fig7}
    \end{figure}

     After the pixels larger than the high threshold are segmented, the pixels smaller than the low threshold in the quantum image should be found and segmented. The low threshold quantum comparator is the same as the high threshold quantum comparator.  Before the low threshold comparison, we reset the high threshold qubits to $\lvert 0\rangle$, so that the low threshold setting can share the three qubits with the high threshold. In addition, the two auxiliary qubits other than the comparison result should be set to $\lvert 0\rangle$, which will be used for  the low threshold comparison. Due to the two comparison results are needed to segment the image pixels between the high and low threshold later,  the two comparison result must be stored separately and saved to the end, so the low threshold comparison result need to be stored in one additional qubit. Other than that, the other parts are the same as the quantum comparator in Fig. \ref{Fig4}(c). If the output result after the comparison by the quantum comparator is $y_L=1$, the pixels in quantum image will be segmented to 000 through the comparison result and the auxiliary qubit, as shown in Fig. \ref{Fig7}.  
    
\begin{figure}
        \centering
        \subfigure[]{\includegraphics[width=6cm]{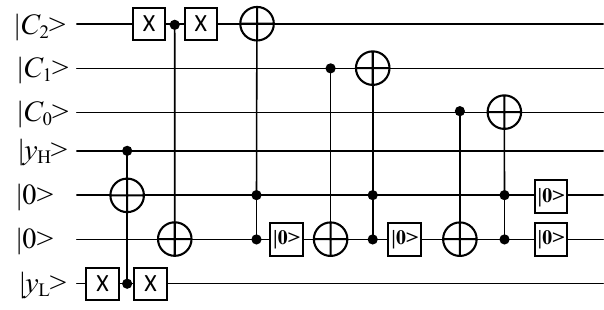 }}
        \subfigure[]{\includegraphics[width=2.5cm]{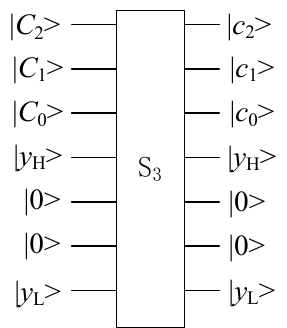 }}
        \caption{The segmentation circuit of the pixel between the high and low thresholds and its simplified diagram $(S_3)$}
        \label{Fig8}
    \end{figure}

\begin{figure}
        \centering
        \includegraphics[width=10cm]{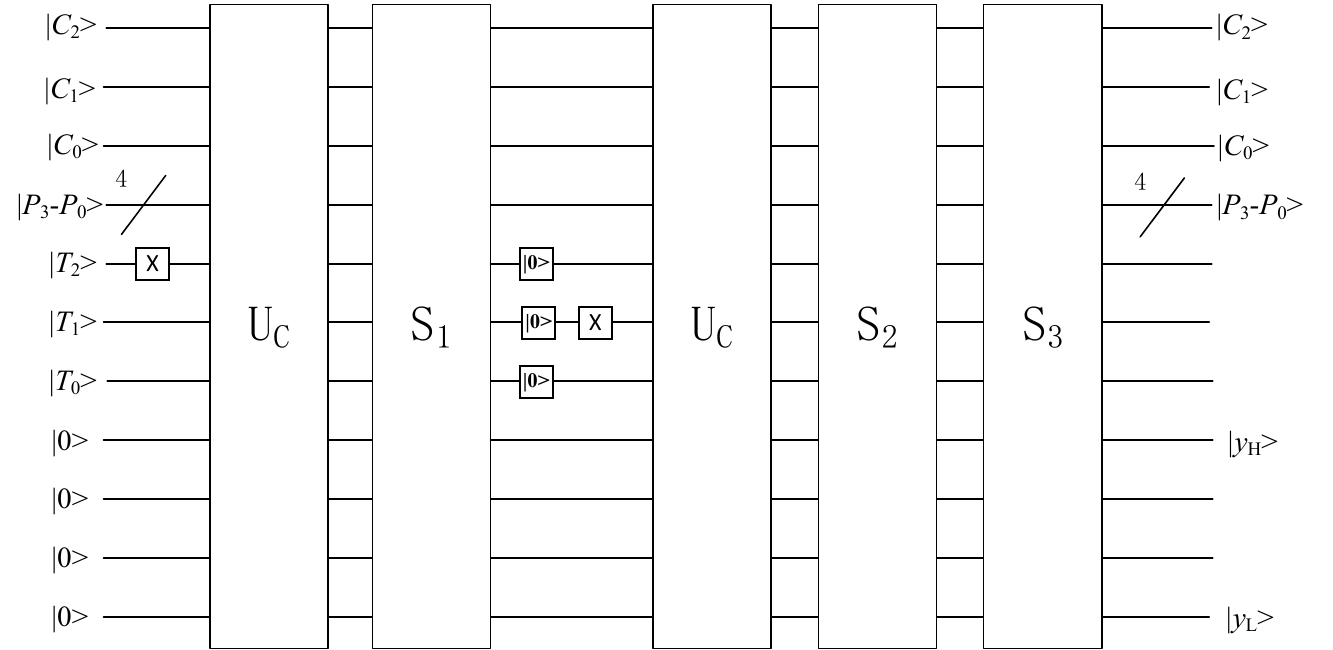 }
        \caption{The complete quantum circuit of quantum image two-threshold segmentation algorithm}
        \label{Fig9}
    \end{figure}

After comparing and segmenting the pixels larger than the high threshold or smaller than the low threshold, the two comparison results are stored in two different qubits. When the pixels gray-scale value in the quantum image are between the high and low threshold, we can find these pixels by using $y_L=0$ and $y_H= 1$. The segmentation method is shown in Fig. \ref{Fig8},  and the whole quantum image segmentation circuit is shown in Fig. \ref{Fig9}.
\section{Circuit quantum cost and experiment analysis}\label{sec4}
In this section,  we  analysis the circuit quantum  cost and compare the performance differences between our algorithm and some other quantum image segmentation algorithms. Then we perform the simulation experiment on the IBM Q platform and the experimental result is analyzed.

\subsection{Circuit quantum  cost analysis}

Single qubit gates and double qubits gates can perform arbitrary operations on any qubits (E.g, NOT gate and CNOT gate), which are basic quantum logic gates.  So, in this paper, we  use  the total number of single qubit gates and double qubits gates (i.e. quantum cost) to evaluate the complexity of quantum circuits. For example, the quantum cost of NOT gate and CNOT gate are both 1. The Toffoli gate can  be decomposed into 5 double qubits gates, thence, the quantum cost of Toffoli gate is 5 \cite{Li2020}. The circuit quantum  cost analysis of the quantum image segmentation algorithm is based on the circuit in Fig. \ref{Fig9}.  Assuming that the size of the quantum image is $2^n\times2^n$, and the gray-scale  range is from 0 to $2^q$-1. The required basic quantum gates  are calculated as follows. Fig. \ref{Fig4} shows that a quantum comparator requires 3$q$-2 Toffoli gates, $q$-1 CNOT gates and 2$q$-2 reset gates, which contains a total of 18$q$-13 basic quantum logic gates, and we need two quantum comparators. It can be seen from Fig. \ref{Fig6}-\ref{Fig8} that the   segmentation operation needs 3$q$+1 Toffoli gates, 3$q$+2 CNOT gates,  and 3$q$+3 reset gates. In addition, 2$q$ NOT gates and $q$ reset gates are also needed to set the threshold. Typically, the NEQR quantum image preparation and measurement processes are not considered parts of quantum image processing. Therefore, the  two-threshold quantum image segmentation algorithm requires a total of 60$q$-6 basic quantum logic gates. So, the circuit quantum cost is 60$q$-6, and the circuit complexity is O($q$),   which means that the complexity is only related to  gray-scale value $q$, and the circuit complexity  will not increase as the  size of the quantum image increases.  
    
Then, the algorithm our proposed is compared with the threshold-based  quantum image segmentation algorithms in references \cite{Caraiman2015,Xia2019,Yuan2020} in terms of the number of threshold, auxiliary qubits and the quantum cost.  For convenience, the segmentation algorithms in \cite{Caraiman2015,Xia2019,Yuan2020} are abbreviated as IS, NMQCIS, DQIS,  respectively. The comparison results are shown in  Tab. \ref{tab2}.

\begin{table}[h]
\begin{center}
\begin{minipage}{320pt}
\caption{Comparison of the different quantum image segmentation algorithms}\label{tab2}%
\resizebox{\textwidth}{16mm}{
\begin{tabular}{@{}lllll@{}}
\toprule
 Algorithms  & Threshold numbers   & Auxiliary qubits  &  Quantum cost &Segmentation numbers\\
\midrule
IS \cite{Caraiman2015}    &  1    & 3$q$-1  & 127$q$-91 &2\\
NMQCIS \cite{Xia2019}    & 1    & 18    & 48$q$-6   &2\\
DQIS \cite{Yuan2020}   & 2      & 5       & 70$q$-14  &2   \\
Our algorithm           &2      & 4        & 60$q$-6   &3 \\ 
\botrule
\end{tabular}}
\end{minipage}
\end{center}
\end{table}
 
  As shown in Tab. \ref{tab2},   one threshold and two QBSCs are used in the IS algorithm \cite{Caraiman2015} to segment the quantum image into a binary image. Scince the quantum cost of 1-qubit QBSC is 14, and the number of auxiliary qubits is 3$q$-1, so the whole circuit quantum cost is 127$q$-91. The NMQCIS algorithm \cite{Xia2019} is also used  to segment the quantum image into a binary one, where one threshold,   one half-comparator, and 2$q$ swap gates with 3 control qubits are introduced. The quantum cost of the half-comparator is 14$q$-6, and    one swap gate  requires 5 Toffoli gates,  so the quantum cost is 48$q$-6. In the DQIS algorithm \cite{Yuan2020}, the author claimed “the number of the required basic gates or operations is $42q + 1$", In fact, the quantum cost is inaccurate because the authors omitted a comparator. In this algorithm, two thresholds and two comparators (the quantum cost of one  comparator  is 28$q$-15) are used  to segment the quantum image, which require 5 auxiliary qubits. And as shown in Fig. 12 in Ref. \cite{Yuan2020}, two comparators,  2$q$+2 Toffoli gates, 2$q$ CNOT gates, 4 NOT gates and 2$q$+2 reset gates are used to compose the complete circuit.  So, the quantum cost of 
 DQIS algorithm is $2(28q-15)+[5(2q+2)]+(2q)+(4)+(2q+2)=70q-14$. In addition, the pixels more than  the high threshold and less than the low threshold are segmented into 0, and the other pixels remain unchanged, which is  inappropriate for complex images. The above algorithms only perform two segmentations, which will result in the inability to segment the details of the image.    It can be seen from Tab. \ref{tab2} that our segmentation algorithm is a two-threshold segmentation algorithm, which has fewer  auxiliary qubits and lower quantum cost  than the other three algorithms.  Then, the image is segmented into a ternary image by three segmentations, which can better segment the complex image and preserve the details in the image. In addition, the circuit quantum  cost will  not  increase as the image's size increase, which is very beneficial for image segmentation with multiple gray-scale levels.

\subsection{Experiment analysis}

Currently, the IBM Quantum Lab provides some tools for running a quantum algorithm. One is the IBM Q  platform, which contains some real quantum computers and simulators on the cloud, and users can upload their own quantum algorithm programs to the cloud to run their own programs on the real quantum computer or simulators.  Another method is to use the open-source quantum computing toolkit Qiskit developed by IBM Research Laboratory to write our own algorithms with the Python language as the front-end interface, so that we can realize the simulation of quantum computers on our own computers, but the the number of qubit is very small.  Therefore, we choose a quantum computer simulator called 'ibmq\_qasm\_simulator' to perform simulation experiments through Qiskit  extension \cite{Qiskit}. This simulator has 32 qubits and can run all the basic logic gates included in our circuit. So in this paper, all quantum circuits have been simulated and verified on this simulator.

\begin{figure}
    \centering
    \subfigure[Schematic  of the original image]{\includegraphics[width=3.5cm]{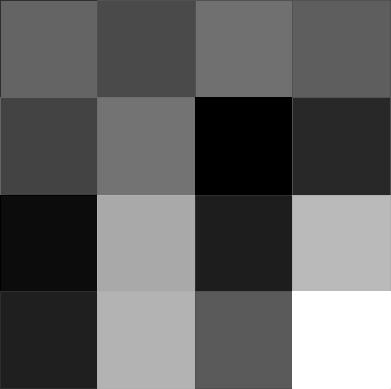}}
    \subfigure[Schematic  of  the segmented image]{\includegraphics[width=3.5cm]{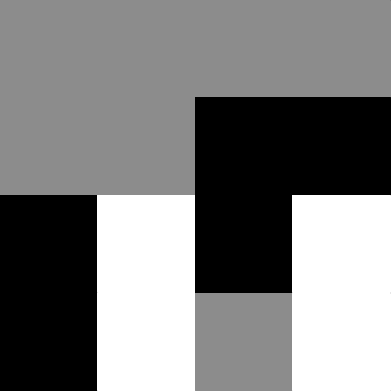}}
    \caption{Schematic  of the original image and the segmented image}
    \label{Fig10}
\end{figure}

To verify the feasibility of the algorithm, we use an image with a size of $4\times4$ and a gray-scale range of [0,7] as an example. Fig. \ref{Fig10}(a) is a schematic diagram of the original image, and Fig. \ref{Fig10}(b) is a schematic diagram of the segmented image. The  NEQR model of the segmented image  is shown in Eq(6).

\[\begin{array}{l}
\lvert {I^{{'}}} \rangle  = \frac{1}{{{2^2}}}(\lvert {1000000} \rangle  + \lvert {1000001} \rangle  + \lvert {1000010} \rangle  + \lvert {1000011} \rangle+\\
\begin{array}{*{20}{c}}
{\begin{array}{*{20}{c}}
{}&{\begin{array}{*{20}{c}}
{}&{}&{}&{}&{}&{}&{}
\end{array}}
\end{array}}&{ \lvert {1000100} \rangle  + }
\end{array} \lvert {1000101} \rangle  + \lvert {0000110} \rangle  + \lvert {0000111}\rangle+\\
\begin{array}{*{20}{c}}
{\begin{array}{*{20}{c}}
{}&{\begin{array}{*{20}{c}}
{}&{}&{}&{}&{}&{}&{}
\end{array}}
\end{array}}&{  \lvert {0001000} \rangle + }
\end{array} \lvert {1111001} \rangle  + \lvert {0001010} \rangle + \lvert {1111011}\rangle+ \\
\begin{array}{*{20}{c}}
{\begin{array}{*{20}{c}}
{}&{\begin{array}{*{20}{c}}
{}&{}&{}&{}&{}&{}&{}
\end{array}}
\end{array}}&{ \lvert {0001100} \rangle + }
\end{array} \lvert {1111101} \rangle  + \lvert {1001110} \rangle + \lvert {1111111} \rangle)\\
\end{array}\eqno{(6)}\]

\begin{figure}
    \centering
    \includegraphics[width=9cm]{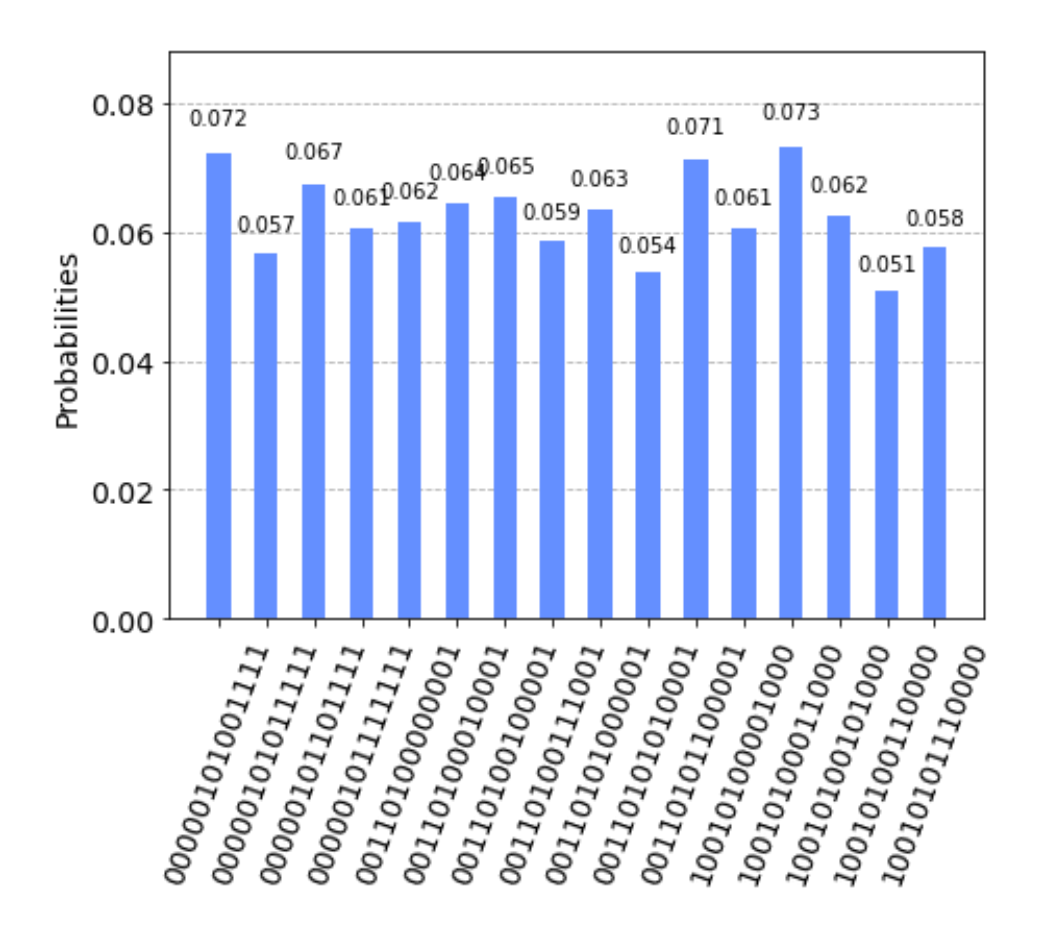}
    \caption{Probability histogram of $4\times4$ segmented quantum image }
    \label{Fig11}
\end{figure}

\begin{figure}
    \centering
  \subfigure[]{ \includegraphics[width=2.5cm]{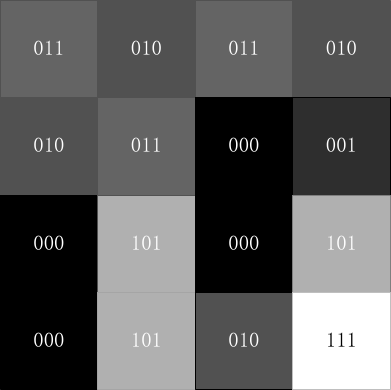 }}
   \subfigure[]{\includegraphics[width=2.5cm]{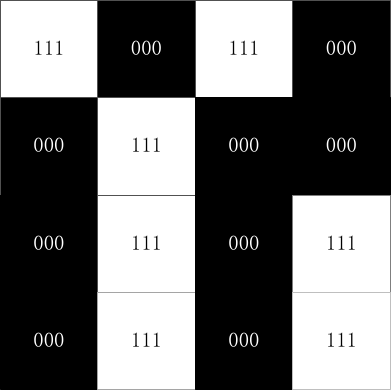 }}
   \subfigure[]{\includegraphics[width=2.5cm]{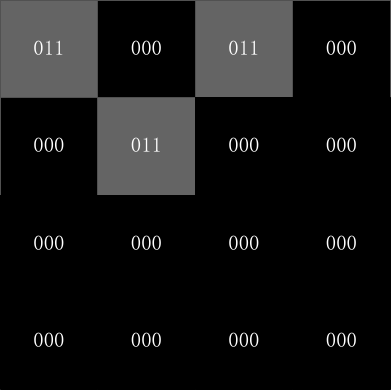 }}
    \subfigure[]{\includegraphics[width=2.5cm]{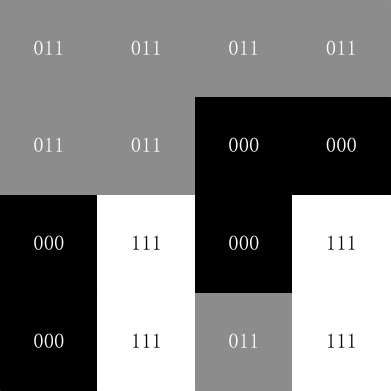 }}\\
   \setcounter{subfigure}{0}\subfigure[]{\includegraphics[width=2.5cm]{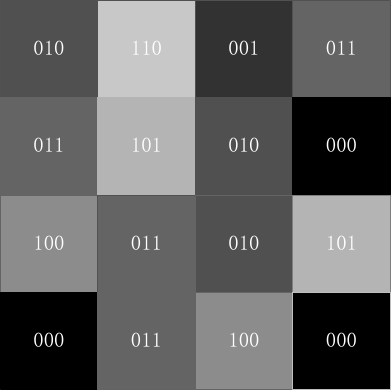 }}
   \subfigure[]{\includegraphics[width=2.5cm]{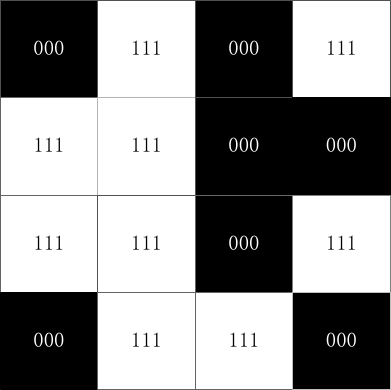 }}
   \subfigure[]{\includegraphics[width=2.5cm]{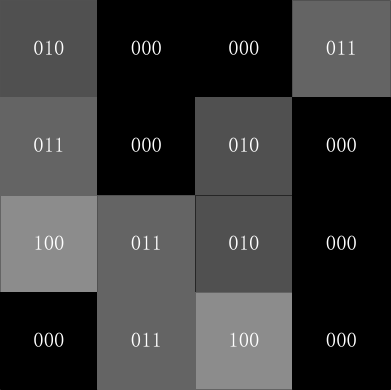 }}
    \subfigure[]{\includegraphics[width=2.5cm]{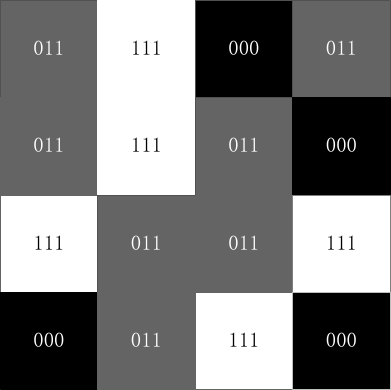 }}\\
   \setcounter{subfigure}{0}\subfigure[]{\includegraphics[width=2.5cm]{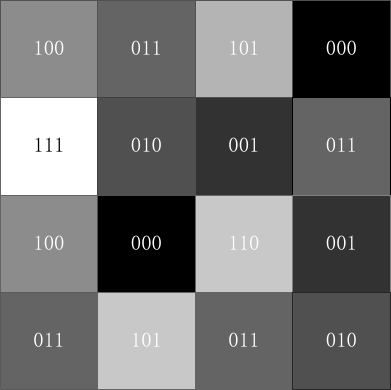}}
   \subfigure[]{\includegraphics[width=2.5cm]{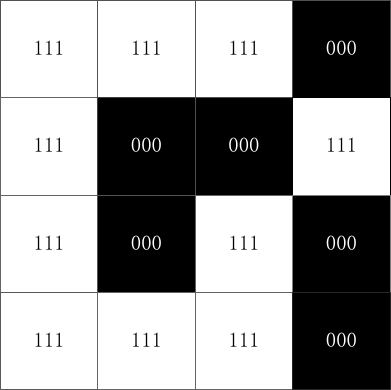}}
   \subfigure[]{\includegraphics[width=2.5cm]{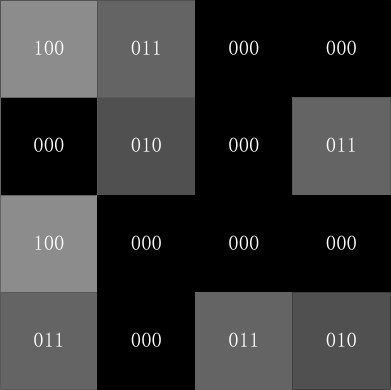 }}
    \subfigure[]{\includegraphics[width=2.5cm]{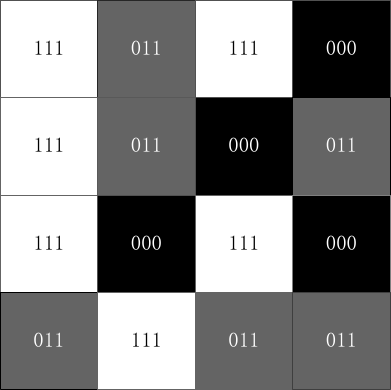 }}\\
    \caption {(a) Three original images. (b) The result images of  the single threshold algorithm(IS and NMQCIS). (c) The result images of  the double threshold algorithm (DQIS).  (d) The result images of our proposed  algorithm. The threshold of (b) is 011; the thresholds of (c) and (d) are $T_H=100$ and $T_L=010$.}
    \label{Fig12}
\end{figure}

Fig. \ref{Fig11}  shows the probability histogram of the segmented image, and  the number of measurements is 1024. It can be seen from the probability histogram that the probability amplitude of the quantum sequence measured is different, which further verifies the randomness of the quantum system and the principle of "uncertainty".  To verify the effect of our algorithm, we compare our algorithm with the existing single threshold (IS and NMQCIS ) and dual-threshold (DQIS) quantum segmentation algorithms, as shown in Fig. \ref{Fig12}. IS and NMQCIS use two segmentations to segment the image into binary images, which is only suitable for simple images and loses a lot of details. Since DQIS also only performs two segmentations, the pixels between the double thresholds are messy, and the rest of the pixels are not classified. This leads to unclear segmentation of the resulting image details. Like the corresponding classical algorithm, our algorithm converts the image into a ternary image with three segmentations, which is very clear and retains more details.

\begin{figure}
    \centering
    \subfigure[$T_L$=011,$T_H$=101 ]{\includegraphics[width=2.5cm]{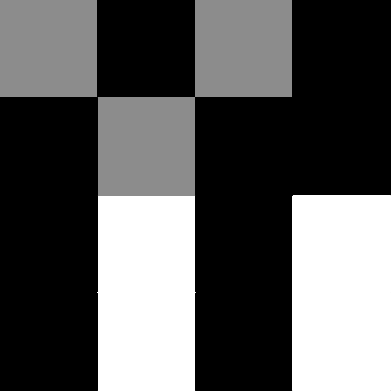}}
    \subfigure[$T_L$=011,$T_H$=110]{\includegraphics[width=2.5cm]{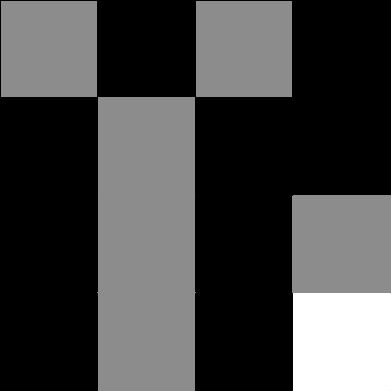}}
    \subfigure[$T_L$=001,$T_H$=011]{\includegraphics[width=2.5cm]{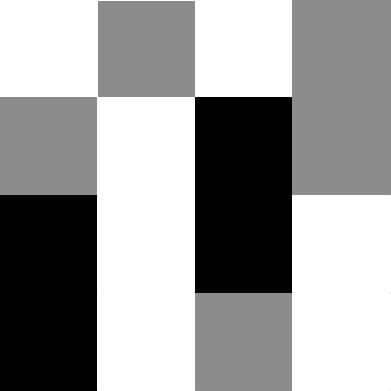}}
    \subfigure[$T_L$=010,$T_H$=101]{\includegraphics[width=2.5cm]{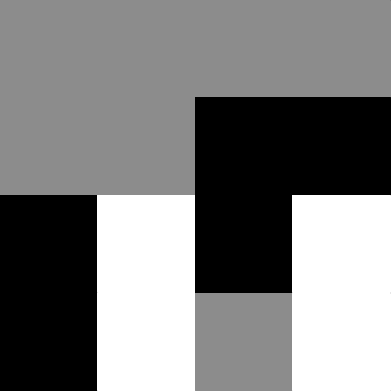}}
    \caption{Segmentation results under different thresholds}
    \label{Fig13}
\end{figure}

Based on the segmentation algorithm in this paper, we also give some schematic diagrams of the segmentation results with different thresholds, as shown in Fig.  \ref{Fig13}.
For quantum image segmentation algorithm simulation, the fewer qubits, basic quantum logic gates, and measurement times are used, the less time it takes, which can directly improve the efficiency of segmentation. In the design of quantum image processing algorithms, there are two ways to directly reduce the quantum cost. One method is to make full use of the parallelism of quantum, which is also an important reason why quantum image processing has exponentially accelerated compared to digital image processing. The other method is to use the reset operation multiple times to reuse the qubits that have been used. In our quantum image segmentation algorithm, we make full use of quantum parallelism and reset operations, so that the quantum  cost is very small and it will not increase as the image size  increases.

\section{Conclusion and discussion}\label{Sec5}

In this paper,  an improved two-threshold quantum  segmentation algorithm for NEQR image is proposed, which can segment the quantum image into a ternary image and can be scaled to use $n$ thresholds for $n+1$ segmentations. In addition, a  quantum comparator and a corresponding quantum segmentation circuit   is designed,  which can segment the quantum image by using few qubits, and the quantum cost  dose not increase as the image's size increases.  The  quantum circuits  analysis and the  experiment on  IBM Q  show that the algorithm can segment  image  efficiently.  However, our algorithm can only segment static targets in static images, which is less effective for dynamic target segmentation. Therefore, our future work is to study the segmentation of dynamic objects in static images. Besides, we will  continue to optimize the quantum circuit and reduce the qubits consumption to increase the applicability of the algorithm.

As described in this paper, our algorithm can divide a gray-scale  image into a ternary image, which can clearly distinguish the objects in the image. However, in some specific cases, it is necessary to obtain an image with more gray-scale values by image segmentation and segment multiple targets in the image. We can scale the proposed algorithm  to solve this problem. To be specific,  the comparators can be extended to n-qubit comparators by increasing the number of comparison qubits, and  we only need to increase the number of comparator for multiple thresholds comparisons. For each additional quantum comparator, we  need to add 
a corresponding segmentation circuit ($S_3$) to segment the pixels between two thresholds. When the previous thresholds segmentation is over, we reset the comparison result qubits so that each additional comparator afterward can reuse these auxiliary qubits. Therefore, we can scale the segmentation algorithm only by adding quantum comparators and segmentation circuits($S_3$). In this way, we can obtained an  image with more gray-scale values. So, our algorithm can use fewer qubits to segment complex images with different gray-scale levels in various situations, which is very beneficial for practical applications. 

\bmhead*{Acknowledgments}
This work is supported by the National Natural Science Foundation of China (62071240),  and the Priority Academic Program Development of Jiangsu Higher Education Institutions (PAPD).

\begin{appendices}
\section{The  circuit on  IBM Q platform}\label{secA1}
    \includegraphics[width=10cm]{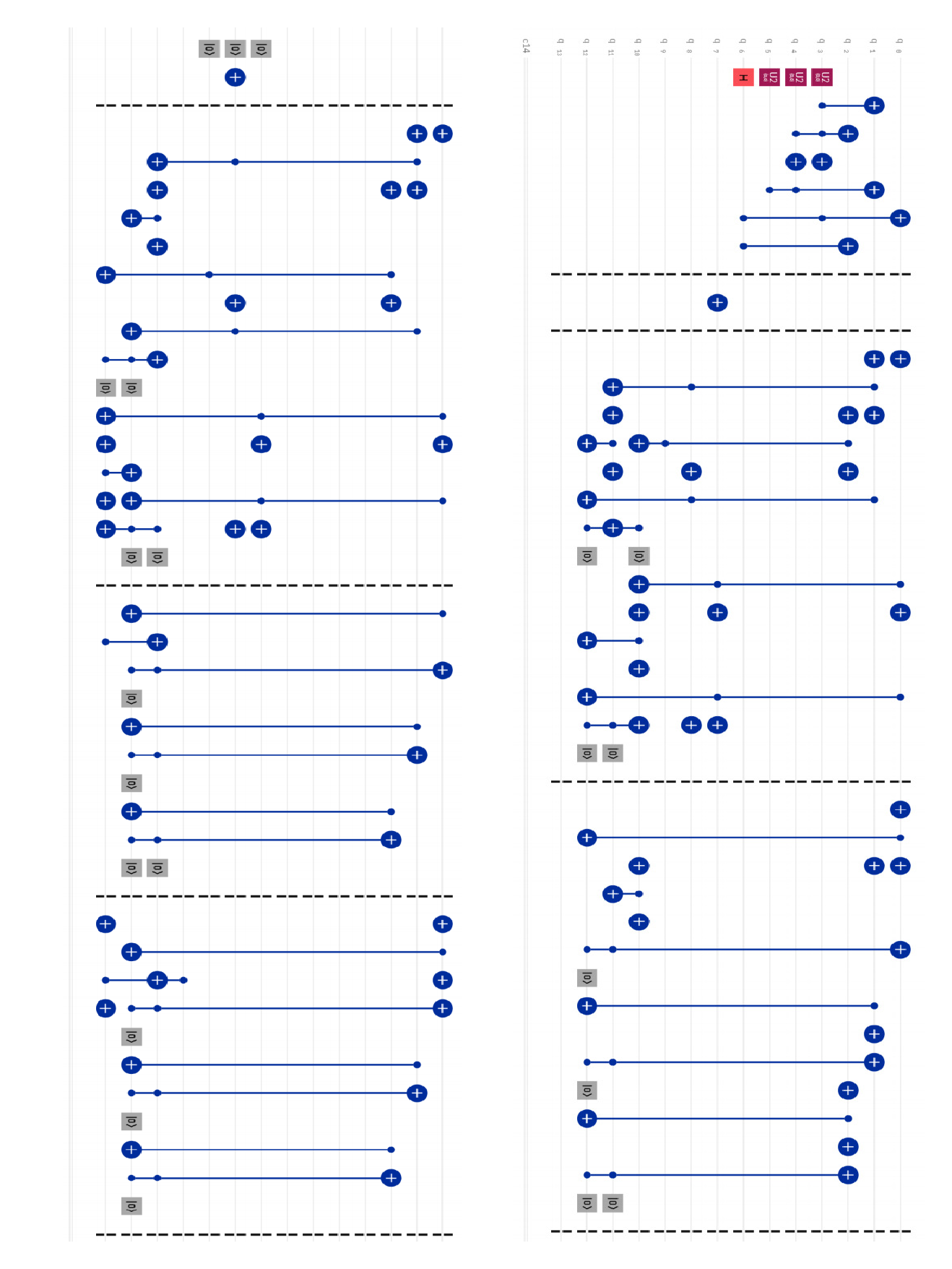 }\\
    Fig.14 The  circuit on IBM Q platform
\end{appendices}

All data generated or analysed during this study are included in this published article [and its supplementary information files].
%%===========================================================================================%%
%% If you are submitting to one of the Nature Portfolio journals, using the eJP submission   %%
%% system, please include the references within the manuscript file itself. You may do this  %%
%% by copying the reference list from your .bbl file, paste it into the main manuscript .tex %%
%% file, and delete the associated \verb+\bibliography+ commands.                            %%
%%===========================================================================================%%

%\bibliography{sn-bibliography}% common bib file
%% if required, the content of .bbl file can be included here once bbl is generated
%%\input sn-article.bbl

%% Default %%
%%\input sn-sample-bib.tex%

\end{document}